\pdfoutput=1 
\documentclass[a4paper, 11pt]{article}
\usepackage[dvipsnames]{xcolor}
\usepackage{jheppub}
\usepackage{physics}
\usepackage{braket}
\usepackage{scalerel}
\usepackage{blindtext} 
\usepackage{epsfig, cancel}
\usepackage{latexsym}
\usepackage{natbib, comment}
\usepackage[all]{xy}
\makeatletter
\gdef\@fpheader{}
\makeatother
\usepackage[T1]{fontenc}
\usepackage{amsmath,amssymb,amsthm,amsfonts,graphicx,hyperref}
\usepackage{stmaryrd}
\usepackage[utf8]{inputenc}
\usepackage{bm}
\hypersetup{ linktoc=all,
    colorlinks, linkcolor=blue,
    citecolor=magenta, urlcolor=red
}

\newcommand{\be}{\begin{equation}}
\newcommand{\ee}{\end{equation}}
\newcommand{\bea}{\begin{eqnarray}}
\newcommand{\eea}{\end{eqnarray}}

\newcommand{\PP}{\text{P}\!\!\!\text{P}}
\newcommand{\PPhi}{\text{P}\!\!\Phi}
\newcommand{\PhiPhi}{\Phi\!\!\!\Phi}

\definecolor{rosy}{RGB}{230,235,252}
\definecolor{myframetitle}{RGB}{90,89,170}
\definecolor{myblocktitle}{RGB}{140,185,249}
\definecolor{mytitle}{RGB}{10,80,26}
\definecolor{darkgreen}{RGB}{27,130,45}
\definecolor{darkblue}{rgb}{0,0,0.3}
\definecolor{darkred}{rgb}{0.7,0,0}
\definecolor{light gray}{RGB}{220,220,220}
\definecolor{dark purple}{RGB}{108,0,217}
\definecolor{pink}{RGB}{190,20,100}
\definecolor{orang}{RGB}{193,63,0}
\definecolor{green}{RGB}{11,98,17}
\definecolor{darkpink}{RGB}{153,0,76}
\definecolor{bluegreen}{RGB}{0,102,102}
\definecolor{greenlagan}{RGB}{0,102,0}
\definecolor{redgreen}{RGB}{102,102,0}
\definecolor{Redgreen}{RGB}{153,76,0}
\definecolor{vividviolet}{rgb}{0.62, 0.0, 1.0}
\definecolor{amaranth}{rgb}{0.9, 0.17, 0.31}
\definecolor{palatinateblue}{rgb}{0.15, 0.23, 0.89}
\definecolor{brightpink}{rgb}{1.0, 0.0, 0.5}
\definecolor{cornflowerblue}{rgb}{0.39, 0.58, 0.93}
\definecolor{deepcarminepink}{rgb}{0.94, 0.19, 0.22}
\definecolor{radicalred}{rgb}{1.0, 0.21, 0.37}
\definecolor{darkmagenta}{rgb}{0.67, 0, 0.67}

\makeatletter \@addtoreset{equation}{section}

\usepackage[most]{tcolorbox}
\tcbset{highlight math style={left=02mm,right=02mm,top=02mm,bottom=02mm}}
\usepackage{empheq}

\begin{document}

\title{{\LARGE{\centerline{Null-strings Gauged, Reloaded and Quantized, I:}}} 
\centerline{\Large{Canonical Quantization in the Light-Cone Gauge}}}    

\author[a]{Ida M. Rasulian}
\author[a]{, M.M. Sheikh-Jabbari}
\author[b]{, H. Yavartanoo}
\affiliation{$^a$ School of Physics, Institute for Research in Fundamental Sciences (IPM), P.O.Box 19395-5531, Tehran, Iran}
\affiliation{$^b$ Beijing Institute of Mathematical Sciences and Applications (BIMSA), Huairou District, Beijing 101408, P. R. China}
\emailAdd{idarasulian@ipm.ir, jabbari@theory.ipm.ac.ir, yavar@bimsa.cn}

\abstract{We study the light-cone quantization of null-strings in $D$ dimensional flat target-space. Incorporating the essential new gauge symmetry and the associated constraint structure of the null-string, allows one to solve for one more degree of freedom (DoF) compared to the standard light-cone gauge, reducing the physical phase space to $(D-3)$ propagating DoF. Quantization is formulated directly in the Schr\"odinger representation, leading to a Hilbert space of wavefunctions on the reduced configuration space.  The space of physical states  is built on a reduced phase space associated with the corrected gauge structure. We discuss the ground-state wavefunction and a generic class of excited states. As a direct consequence of the overlooked Carroll-Weyl gauge symmetry of the null-strings \cite{Sheikh-Jabbari:2026cnj} we find the remarkable, and perhaps unexpected, result that null-strings exhibit a discrete spectrum. Our analysis indicate that there is no critical dimension for null-strings, and $D$ can be arbitrary.}  
\maketitle
\flushbottom

\section{Introduction}

Null-strings occupy a special position among string theories. They constitute a class of tensionless strings and may be obtained as the zero-tension limit of ordinary strings, nonetheless, their properties differ substantially from those of their tensile counterparts; see the introduction in \cite{Sheikh-Jabbari:2026vqh} for tensionless vs null-string  discussion. In particular, the vanishing of the string tension removes the conventional oscillator dynamics that underlies much of the structure of the ordinary bosonic string. As a result, many familiar features of string quantization must be reconsidered from first principles.

The study of null-strings has a long history; see \cite{Bagchi:2026wcu} and references therein. Early formulations were developed by Schild \cite{Schild:1976vq}, and subsequently by Isberg-Lindstrom-Sundborg-Theodoridis (ILST) \cite{Lindstrom:1990qb, Isberg:1992ia, Isberg:1993av}. Afterward, many works have investigated  classical and quantum aspects of null-strings both as independent dynamical systems and as the tensionless limit  of tensile string theory \cite{Gustafsson:1994kr, Jensen:1996dc, Bagchi:2020fpr, Bagchi:2021rfw, Bagchi:2019cay, Bagchi:2022iqb, Banerjee:2024fbi, Figueroa-OFarrill:2025njv}. The worldsheet of a null-string  is a $2d$ null surface, it is a two-dimensional Carrollian geometry. As such, null-strings naturally arise when we deal with a relativistic gas of strings close to Hagedorn temperature \cite{Hagedorn:1965st, Lizzi:1986nv}, or when strings probe regions of the target-space  with large redshift factors, when strings probe cosmological or black hole horizons \cite{Bagchi:2021ban,Bagchi:2023cfp, Bagchi:2024rje, Bagchi:2026wcu}. 

Recently, it was noted that in the null-string  analysis  a crucial point has been overlooked \cite{Sheikh-Jabbari:2026cnj}: The ILST action possesses an additional gauge symmetry, which was dubbed as Carroll-Weyl gauge symmetry \cite{Sheikh-Jabbari:2026vqh}. This gauge symmetry has no counterpart in the tensile string and only appears on Carrollian worldsheets due to specific features of Carrollian geometry \cite{Sheikh-Jabbari:2026vqh}. Due to the presence of this extra gauge symmetry, one more of the string embedding fields becomes non-physical gauge degree of freedom (DoF).  This gauge symmetry has been discussed and established at classical level in the Hamiltonian formulation in the light-cone gauge \cite{Sheikh-Jabbari:2026tpf}. The identification of the Carroll-Weyl gauge symmetry calls into question all  null-string  treatments prior to \cite{Sheikh-Jabbari:2026cnj,Sheikh-Jabbari:2026vqh, Sheikh-Jabbari:2026tpf} and has prompted a new reanalysis of null-string  literature \cite{Lindstrom:2026quz, Duary:2026rlo, Lindstrom:2026zno}.  

The realization of the extra Carroll-Weyl gauge symmetry calls into a serious and thorough revision of not only the classical treatments of null-strings, but also its quantization. There are various ways of quantizing a gauge theory or string theory \cite{Green:1987sp, Polchinski:1998rq}: canonical quantization which may be carried out covariantly or in the light-cone gauge, path integral quantization and introducing ghosts. In the canonical quantization there is also a third, less known and common, method: quantization in Schr\"odinger representation, in which physical states are wavefunctionals on the reduced configuration space \cite{Hopkinson:1975pm, Kanatchikov:2000yh, Hatfield:2019sox}. Quantization in the  Schr\"odinger representation for strings is somewhat halfway through the ordinary quantization of worldsheet theory and the string field theory.  Our choice of the Schr\"odinger representation is motivated by the 
fact that, unlike the ordinary tensile bosonic string whose reduced Hamiltonian is a sum of harmonic oscillators and naturally admits a Fock-space construction, the reduced null-string Hamiltonian is not of harmonic-oscillator type. Consequently, the Schr\"odinger representation provides a more natural starting point for quantization. One of the aims of this paper is to show that this representation provides a natural framework for  a consistent quantization of null-string theory.

We carry  the  out quantization in the Schr\"odinger representation in the ``null-string  light-cone gauge'' (NSLCG), which has specific features different from ordinary tensile string light-cone gauge 
\cite{Sheikh-Jabbari:2026tpf}. In this work we focus on null-strings on a $D$-dimensional flat target-space and show that it has $D-3$ propagating modes, in contrast with $D-2$ of ordinary tensile strings in the light-cone gauge \cite{Sheikh-Jabbari:2026cnj}; these $D-3$ modes  are subject to  a ``null-string  level-matching'' condition. Our analysis leads to two fundamental results: (1) within the present framework we find no evidence for a critical dimension, and the target-space dimension $D$ remains unrestricted;\footnote{There are conflicting statements on the critical dimensions in the null-string  literature,  see e.g. \cite{Lizzi:1986nv, Isberg:1993av, Gustafsson:1994kr, Murase:2015yaa, Bagchi:2021rfw, Bagchi:2026wcu, Chen:2026klv}.} (2) the physical spectrum is discrete. Both conclusions differ qualitatively from the ordinary tensile string, where the critical dimension and the infinite tower of massive string states are direct consequences of string tension. In the tensionless null-string  case, our results arise as a consequence of the additional Carroll-Weyl gauge symmetry.

The paper is organized as follows. Section~\ref{sec:ClassicalReview} summarizes the classical ingredients required for quantization, in particular NSLCG fixing. Section~\ref{sec:quantization} we work through  quantization in Schr\"odinger representation  and constructs  the vacuum (ground state) of the null-string  theory. In Section~\ref{sec:quadratic-states} we study a class of excited null-string  states and work out the mass spectrum and light-cone Hamiltonian wavefunctional. We explicitly demonstrate quantization and discreteness of null-string  excitations masses in this sector. We close in section \ref{sec:Outlook} by concluding remarks and outlook.  Appendix~\ref{app:Tensile} contains a review of the tensile-string Schr\"odinger representation used as a point of comparison throughout the paper.

\section{Classical Light-Cone Gauge Reduction and Symplectic Geometry}\label{sec:ClassicalReview}

The purpose of this section is to set the stage for quantization of null-strings. This is a review of 
\cite{Sheikh-Jabbari:2026cnj, Sheikh-Jabbari:2026tpf, Sheikh-Jabbari:2026vqh}. We carefully go through fixing the null-string  light-cone gauge (NSLCG) and solve the three constraints resulting from the gauge symmetries of the ILST \cite{Isberg:1993av} null-string  worldsheet theory.

\subsection{Classical action and  constraint structure}

We begin with the ILST action 
\begin{equation}\label{ILST-Action}
S = -\frac{\kappa}{2} \int d\tau d\sigma\, \mathcal V^a \mathcal V^b \partial_aX^\mu \partial_bX_\mu ,
\end{equation}
where ${\cal V}^a$ is a non-dynamical field that encodes the kernel vector of the $2d$ Carrollian 
geometry \cite{Sheikh-Jabbari:2026vqh} and $\sigma^a=(\tau, \sigma)$ denote the worldsheet coordinates. This action is invariant under $2d$ diffeomorphisms generated by $\xi^a=\xi^a(\tau, \sigma)$ and, as importantly noted in \cite{Sheikh-Jabbari:2026cnj},  also exhibits a gauge invariant under a novel symmetry generated by $\chi=\chi(\sigma)$.\footnote{It was discussed in \cite{Sheikh-Jabbari:2026vqh} that one can extend the $\chi(\sigma)$ symmetry to a fully fledged gauge symmetry generated by $\chi(\tau, \sigma)$, recalling the structures natural to Carrollian geometry of the worldsheet. This extension entails introducing an auxiliary `gauge field' ${\cal W}_a$ and replacing $\partial_a$ with a covariant derivative $D_a$, $D_aX^\mu = \partial_aX^\mu + \mathcal W_a X^\mu$. Upon fixing the $\chi$-symmetry (Carroll-Weyl symmetry), the gauged action reduces to ILST action \eqref{ILST-Action} and $\chi$ reduces to $\chi=\chi(\sigma)$. Therefore, the dynamical content of the gauged action and the ILST are exactly the same. See \cite{Sheikh-Jabbari:2026tpf, Sheikh-Jabbari:2026vqh} for detailed analysis. So, we restrict our analysis to the ILST action and do not discuss the gauged action. } If we denote these two gauge symmetry parameters collectively as $\eta=(\xi^a, \chi)$,
\begin{equation}\label{eta-transform}
\begin{split}
\delta_\eta X^\mu&= \xi^a\partial_a X^\mu + \chi X^\mu\\
  \delta_\eta {\cal V}^a &= \xi^b\partial_b{\cal V}^a - {\cal V}^b\partial_b\xi^a  +(\frac12\partial_b\xi^b-\chi) {\cal V}^a
\end{split}\end{equation}
One can leverage the diffeomorphism invariance to fix temporal gauge,
\begin{equation}
\mathcal V^a = (1,0),
\end{equation}
leaving us with a residual symmetry of $\tau$-independent diffeomorphisms. Hence the residual symmetries are generated by three functions of $\sigma$, $\eta=(\zeta^a(\sigma), \chi(\sigma))$, with
\begin{equation}\label{residual-zeta}
    \zeta^a= [h(\sigma)+ (f'(\sigma) - 2\chi(\sigma))\tau]\partial_\tau+ f(\sigma)\partial_\sigma\,,\qquad f'=\partial_\sigma f. 
\end{equation}
{Associated with these three residual gauge symmetries are three constraints; two of them correspond to the equations of motion (EoM) for ${\cal V}^a$, denoted by ${\cal C}_1$ and ${\cal C}_2$, while the third, ${\cal C}_3$, is the constraint associated with the $\chi(\sigma)$ symmetry \cite{Sheikh-Jabbari:2026cnj, Sheikh-Jabbari:2026tpf, Sheikh-Jabbari:2026vqh}:}
\begin{align}
\mathcal C_1 &= P^2 \approx 0 , \label{C1} \\
\mathcal C_2 &= P\cdot X' \approx 0 , \label{C2} \\
\mathcal C_3 &= P\cdot X \approx 0, \label{C3}
\end{align}
where $X'=\partial_\sigma X$ and $P$ is the canonical momentum to $X^\mu$,
\begin{equation}
P^\mu := \kappa \mathcal V^a \partial_aX^\mu =\kappa \partial_\tau X^\mu:= \kappa \dot{X}^\mu.
\label{CanonicalMomentum}
\end{equation}

The EoM in the temporal gauge take the simple form
\begin{equation}
\dot P^\mu = 0  \qquad\text{or}\qquad  \ddot X^\mu =0 .
\end{equation}
The general solution is therefore given by:
\begin{equation}\label{GeneralSolution}
X^\mu(\tau,\sigma) = X_0^\mu(\sigma) + \frac{1}{\kappa} P^\mu(\sigma) \tau .
\end{equation}
Thus, the classical dynamics is completely determined by two $\sigma$-dependent fields, $X_0^\mu(\sigma)$ and $P^\mu(\sigma)$. Substituting the general solution \eqref{GeneralSolution} into constraints \eqref{C1}--\eqref{C3} maps the originally time-dependent system into a set of purely spatial conditions:
\begin{align}
\texttt{C}_1&:=P^2(\sigma) = 0 , \label{ConstraintP2} \\
\texttt{C}_2&:=P(\sigma) \cdot X_0'(\sigma) = 0 , \label{ConstraintPXPrime} \\
\texttt{C}_3&:= P(\sigma) \cdot X_0(\sigma) = 0 . \label{ConstraintPX}
\end{align}
Consequently, the classical physical null-string  configurations are specified by  $X_0^\mu(\sigma)$ and $P^\mu(\sigma)$, subject to constraints $\texttt{C}_1$, $\texttt{C}_2$, and $\texttt{C}_3$, up to the residual gauge symmetry identifications \cite{Sheikh-Jabbari:2026vqh}:
\begin{equation}\label{residual-identifications}
\delta_\eta X_0^\mu= f X'_0{}^\mu+ \chi X^\mu_0, \qquad \delta_\eta P_\mu = f P'_\mu+ (f'-\chi) P_\mu.
\end{equation}
{As established in \cite{Sheikh-Jabbari:2026cnj, Sheikh-Jabbari:2026vqh}, out of  $2D$ variables $X_0^\mu(\sigma)$ and $P^\mu(\sigma)$, imposing three constraints and implementing three  gauge-identifications reduces the physical system to $2(D-3)$ independent degrees of freedom.}
\subsection{Null-string light-cone gauge fixing}
To isolate the $2(D-3)$ independent physical variables explicitly, we adopt the null-string  light-cone gauge (NSLCG). We introduce standard light-cone coordinates,
\begin{equation}
X^\pm = \frac{1}{\sqrt{2}}\left(X^0\pm X^{D-1}\right),
\end{equation}
and similarly for the momenta. Throughout the remainder of this paper, we decompose the target-space  coordinates according to 
\begin{equation}
X^\mu = \left(X^+,X^-,X^1,X^I\right),
\qquad
I=2,\ldots,D-2.
\end{equation}
The distinguished coordinate $X^1$  is utilized to fix the residual gauge freedom associated with the final constraint, while the remaining components $X^I$ span the physical transverse sector. The Minkowski inner product takes the form
\begin{equation}
A\cdot B = -A^+B^- - A^-B^+ + A^1B^1 + A^IB^I.
\end{equation}

{We exploit the residual gauge transformations parameterized by $\chi=f'$ and $h=0$ (cf. \eqref{residual-zeta}) to set:}
\begin{equation}
X^+ = \frac{\bar{p}{}^+}{\kappa}\tau,
\qquad
{P}^+ = \bar{p}{}^+,
\qquad
\bar{p}^+\neq0.
\label{eq:LCGaugeReview}
\end{equation}
The constant momentum $\bar{p}^+$ serves as a fixed longitudinal momentum density and allows the constraints to be solved algebraically. {Under this gauge choice, the residual transformations simplify to $\delta_\eta X^\mu = (fX_0^\mu)'$ and $\delta P_\mu = f P'_\mu$.}

{We first evaluate the $\texttt{C}_1=0$ constraint \eqref{ConstraintP2} upon the  gauge condition \eqref{eq:LCGaugeReview} and solve  for $P^-$:}
\begin{equation}
\boxed{\ P^- = \frac{(P^1)^2 + P^I P^I}{2\bar{p}^+}\ }
\label{PMinusSolution}
\end{equation}
{Thus, $P^-$ is not an independent degree of freedom, but is completely determined by the transverse momenta and the remaining longitudinal component ${p}^1$.}

{Next, the $\texttt{C}_3=0$ constraint \eqref{ConstraintPX} yields:}
\begin{equation}
\boxed{\ X_0^- = \frac{{P}^1 {X_0}^1 + P^I X_0^I}{\bar{p}^+}\ }
\label{XMinusSolution}
\end{equation}
{In contrast to the tensile string theory, the coordinate $X_0^-$ is specified through a purely algebraic relation, requiring no integration of the worldsheet constraints.}

The remaining constraint $\texttt{C}_2 =0 $ \eqref{ConstraintPXPrime} takes the form
\begin{equation}\label{C2-NSLCG}
P^{1\prime} {X_0}^1 + P^{I\prime} X_0^I = 0.
\end{equation}
{We utilize the final residual gauge freedom $f$ to set:}
\begin{equation}
{X_0}^1(\sigma) = \bar{x}^1,
\qquad
\bar{x}^1 \neq 0.
\label{eq:X1GaugeReview}
\end{equation}
{Following this gauge-fixing step, the residual gauge symmetries are reduced to the $f = \text{const.}$ sector, which corresponds to rigid translations in $\sigma$. Under this condition, the constraint \eqref{C2-NSLCG} produces a differential equation for ${P}^1$:}
\begin{equation}
P^{1\prime} = -\frac{1}{\bar{x}^1} P^{I\prime} X_0^I.
\label{eq:P1Prime}
\end{equation}
Integrating this relation along the spatial worldsheet circle yields
\begin{equation}\label{P1Solution-f0-not-fixed}
\ {P}^1(\sigma) =  {P}^1(\sigma_0)- \frac{1}{\bar{x}^1} \int_{\sigma_0}^{\sigma} d\bar\sigma\, P^{I\prime}(\bar\sigma) X_0^I(\bar\sigma)\ 
\end{equation}
where $\sigma_0$ is an arbitrary value in $[0,2\pi]$ range. Under the residual rigid $\sigma$-translations $f_0$, $\delta_{f_0} {P}^1(\sigma_0)= - \frac{f_0}{\bar{x}^1}P^{I\prime}(\sigma_0) X_0^I(\sigma_0)$. If we choose $\sigma_0$ to be a root of $P^{I\prime}X_0^I$ then ${P}^1(\sigma_0)$ becomes a fixed constant. One can use $f_0$ to set $\sigma_0=0$. Therefore, all in all, one can  set the $P^1(\sigma_0)=\bar{p}_1$ which is, like $\bar{x}^1, \bar{p}^+$, a given constant, and write 
\begin{equation}\label{P1Solution}
{P}^1(\sigma) =  \bar{p}^1- \frac{1}{\bar{x}^1} \int_{0}^{\sigma} d\bar\sigma\, P^{I\prime}(\bar\sigma) X_0^I(\bar\sigma)\ 
\end{equation}
and all the residual gauge symmetries have been used and all constraints are solved. 

For a closed strings with zero winding number,  ${P}^1(\sigma)$ must remain strictly single-valued under the periodic translation $\sigma \to \sigma + 2\pi$. Consequently, its derivative profile must integrate to zero across the closed loop:
\begin{equation}\label{eq:global_integrability}
\int_0^{2\pi} d\sigma\, P^{1\prime}(\sigma) = 0 \implies \int_0^{2\pi} d\sigma\, P^{I\prime}(\sigma)X_0^I(\sigma) = 0. 
\end{equation}

At this stage, all longitudinal variables are expressed in terms of the $2(D-3)$ transverse coordinates $X_0^I$ and momenta $P^I$ ($I=2,\ldots,D-2$), alongside the three zero-mode parameters $\bar{p}^+, \bar{p}^1$ and $\bar{x}^1$. A NSLCG sector is hence specified by these three parameters. As demonstrated, $X^-_0, P^-$, and ${P}^1$ are explicitly determined in terms of these independent physical variables.

\subsection{Fourier mode expansions and the residual global constraint}

Hereafter, we restrict our attention to closed strings with zero winding, for which all physical fields must be $2\pi$-periodic functions of $\sigma$. They therefore admit standard discrete mode expansions:
\begin{align}\label{eq:XMode_rev}
P^I(\sigma) = p^I + \sum_{n\neq0} P_n^I e^{in\sigma}, \qquad 
X_0^I(\sigma) = x_0^I + \sum_{n\neq0} X_n^I e^{in\sigma}, 
\end{align}
where  reality of the target-space fields requires $(P_n^I)^*=P_{-n}^I$ and $(X_n^I)^*=X_{-n}^I$. Recalling \eqref{P1Solution}, the modes of $P_1(\sigma)$  can then be written in terms of the $X^I_n, P^I_n$ modes:
\begin{equation}\label{eq:P1ModeExact}
\begin{split}
    {P}^1(\sigma) =& \bar{p}^1 + \sum_{n\neq0} P_n^1 e^{in\sigma},\\
P_n^1 = -\frac{x_0^I}{\bar{x}^1} P_n^I - \frac{1}{\bar{x}^1} \Phi_n, & \qquad \Phi_n := \sum_{r\neq0, n} \frac{r}{n} P_r^I X_{n-r}^I, \qquad n\neq0.
\end{split}
\end{equation}
The global integrability condition \eqref{eq:global_integrability} isolates a single global constraint mode:
\begin{equation}\label{eq:G0Review}
\boxed{\ G_0 := \sum_{n\neq0} nP_n^I X_{-n}^I = 0\ } 
\end{equation}
The relation $G_0=0$ is not an additional constraint imposed by hand; rather, it is the constraint that is needed for the consistency of the discussions below \eqref{P1Solution-f0-not-fixed} to allow for fixing $\bar{p}^1$. Thus, \eqref{eq:G0Review} ensures that no preferred point exists on the constant time-slice of the null-string. We refer to \eqref{eq:G0Review} as the ``null-string  level-matching condition''. This condition serves as the null analog of the standard level-matching condition in tensile string theory \cite{Green:1987sp, Polchinski:1998rq}, with the noteworthy distinction that the index $I$ spans $D-3$ transverse directions rather than the $D-2$ directions of the tensile framework.

\paragraph{Light-cone Hamiltonian.} As in the standard light-cone gauge formulation for tensile strings, one can now express the longitudinal coordinate modes $X^-$ and the light-cone Hamiltonian $P^-$ in terms of the transverse modes:
\begin{equation}\label{Phi-Xi-n}
\begin{split}
X_n^- = \frac{1}{\bar{p}^+}  \sum_{r\neq 0} \frac{n-r}{n} P_{r}^I X_{n-r}^I  \quad (n\neq 0),  \qquad X_0^- = \frac{1}{\bar{p}^+}\big(\bar{p}^1 \bar{x}^1+ p^I x_0^I+ \sum_{r\neq 0} P^I_r X^I_{-r}\big)
\end{split}
\end{equation}
\begin{equation}\label{eq:Hamiltonian_Full_Rigorous}
P^-:=H_{\text{LC}} = \frac{1}{2\bar{p}^+} \left[ (\bar{p}^1)^2 + p^Ip^I + {\cal M}^2\right],\qquad {\cal M}^2:=\sum_{n\neq0} \left( P_{-n}^1P_n^1 + P_{-n}^IP_n^I \right). 
\end{equation}
Using \eqref{eq:P1ModeExact}, we have
\begin{equation}
{\cal M}^2= {\cal G}_{IJ}\ {\text{P}\!\!\!\text{P}}^{IJ} + \frac{2 x_0^I}{(\bar{x}^1)^2} \text{P}\!\!\Phi_I + \frac{1}{(\bar{x}^1)^2} \Phi\!\!\!\Phi
\end{equation}
with 
\begin{equation}
{\cal G}^{IJ} := \delta^{IJ}+ \frac{x_0^I x_0^J}{(\bar{x}^1)^2}, \qquad {\PP}^{IJ}:=\sum_{n\neq0} P_{-n}^I P_n^J,\qquad \PPhi_I:=\sum_{n\neq0}  P_{-n}^I \Phi_n,\qquad   \PhiPhi:=\sum_{n\neq0} \Phi_{-n} \Phi_n .
\end{equation}

We pause for some comments on the light-cone Hamiltonian $H_{\text{LC}}$ \eqref{eq:Hamiltonian_Full_Rigorous}.
\begin{itemize}
\item As the notation suggests, ${\cal M}^2$ may be viewed as an effective mass of the null-string  state  from the target-space  viewpoint. 
    \item $\Phi_n$ are Fourier modes of a field which involves integral of products of $P', X$, $\Phi(\sigma)\sim \int^\sigma d\tilde\sigma (P'\cdot X_0)(\tilde\sigma)$ (cf. \eqref{P1Solution} and \eqref{eq:P1ModeExact}). Thus,  $H_{\text{LC}}$ has a highly non-local expression once written in terms of fields $X_0^I(\sigma), P^I(\sigma)$.  Recall also the Carrollian nature of the worldsheet and that no points of the null-string  on a given $\tau$ slice are in causal contact.
    \item Null-strings consists of a congruence of null geodesics in the target-space  where $\tau$ is the affine parameter on the null rays and $\sigma$ parametrizes different rays in the congruence. Of course,  this congruence becomes a null-string  only after imposing the three constraints $\texttt{C}_1, \texttt{C}_2, \texttt{C}_3$. The  non-locality in $H_{\text{LC}}$ is then a direct consequence of these constraints, once the null-string  is explicitly written in terms of the $2(D-3)$ ``transverse modes''. 
\item Unlike the light-cone Hamiltonian of tensile strings on flat space background (which is quadratic in $X$ or $P$), $H_{\text{LC}}$ involves $\PP$ which is quadratic in momenta, the cubic non-local term $\PPhi$ of the form $\sim \int d\sigma\ x_0\cdot P(\sigma) \int^\sigma d\tilde\sigma (P'\cdot X_0)(\tilde\sigma)$ and the quartic non-local term $\PhiPhi$  of the form $\sim \int d\sigma \int^\sigma d\sigma_1 (P'\cdot X_0)(\sigma_1) \int^\sigma d\sigma_2 (P'\cdot X_0)(\sigma_2)$. The non-local terms come with factors of $1/(\bar{x}^1)^2$ which, as we will see, plays the role of an effective tension for the null-string. In other words, ``effective null-string  tension'' is not a parameter of the (ILST) action, but a parameter of null-string  solutions.
\item ${\cal M}^2$ is also depending on $x_0^I$; the first term has quadratic dependence, the middle term linear and the last term has no $x_0^I$ dependence. 
\end{itemize}
{In summary, the null-string  theory in the NSLCG is completely described by three parameters $\bar{p}^+,\bar{p}^1, \bar{x}^1$ and, the $(D-3)$ transverse coordinates and their conjugate momenta, subject to the null-string level-matching condition \eqref{eq:G0Review}.}
\subsection{Reduced phase space }

{To prepare for canonical quantization, we define the classical Poisson brackets on the solution-space fields $X_0^I(\sigma)$ and $P^J(\sigma')$. The equal-time Poisson brackets on  $X^I(\sigma, \tau), P^I(\sigma, \tau)$ yield:}
\begin{equation}
\left\{X_0^I(\sigma), P^J(\sigma')\right\} = \delta^{IJ} \delta(\sigma-\sigma').
\label{eq:ReducedFieldBracket}
\end{equation}
Expanding the Dirac delta function in in term of Fourier modes,
\begin{equation}
\delta(\sigma-\sigma') = \frac{1}{2\pi} \sum_{n\in\mathbb Z} e^{in(\sigma-\sigma')},
\end{equation}
leads to the corresponding algebra for the mode coefficients,
\begin{equation}\label{eq:OscillatoryModePB}
\{x_0^I, p^J\} = \delta^{IJ},\qquad {\{X_n^I, P_m^J\} = \delta^{IJ} \delta_{n+m,0}}.
\end{equation}
All other fundamental Poisson brackets vanish identically. As noted previously, the physical configuration space is constrained by \eqref{eq:G0Review}, and it is straightforward to verify that the physical phase space is closed under this condition. These relations fully define the reduced phase space of the theory, serving as the foundation for the quantization procedure detailed in the subsequent section.

We conclude this section by noting the following helpful bracket relations: 
\begin{equation}\label{G0--H-comm}
\begin{split}
 \{G_0,X_n^I\} = nX_n^I, \qquad  \{ G_0, H_{\text{LC}}&\} =0 ,\qquad \{G_0,P_n^I\} = nP_n^I, \\
      \{ G_0, \Phi_n\} =n\Phi_n,\qquad\   \{ \Phi_m, \Phi_n\} &=\frac{m^2-n^2}{mn}\Phi_{m+n},\quad m,n\neq 0
\end{split}
\end{equation}
In particular, these relations imply that $\{ \Phi_{-n}, \Phi_n\}=0$.

\section{Schr\"odinger Quantization of Null-String  Theory}\label{sec:quantization}

So far, we have constructed the classical solution phase space and we  quantized the system by promoting solution variables, $X_n^I, P_n^I\ (n\neq 0)$ and $x_0^I, p^I$ into operators and Poisson brackets into commutators,\footnote{Note that $\bar{p}^+, \bar{x}^1, \bar{p}^1$ are solution parameters and should be treated as numbers and not operators.}
\begin{equation}
[X_n^I, P_m^J] = i \,\delta^{IJ}\delta_{n+m,0},\qquad [x_0^I, p^J]=i \,\delta^{IJ}.
\end{equation}
The reality conditions inherited from the classical theory imply the Hermiticity relations
\begin{equation}\label{XP-dagger}
(X_n^I)^\dagger = X_{-n}^I, \qquad (P_n^I)^\dagger = P_{-n}^I.
\end{equation}
In doing so, one should mind possible ordering issues that may arise. In particular we note that since for $n \neq 0$, the operators $P_r^I$ and $X_{n-r}^I$ commute, there is no ordering ambiguity in $\Phi_n$. In a similar way, since for $n\neq 0$, $[x_0^I, P^J_n]=0$, $[P_n^I, \Phi_{-n}]=0$, the $\PPhi$ term is free of ordering ambiguity. Despite the fact that $[\Phi_n, \Phi_{-n}]=0$, being a quartic term, $\PhiPhi$ has an ordering issue that we will discuss below. 

Dealing with a non-quadratic and non-local light-cone Hamiltonian, it is not possible to proceed with the canonical quantization as we do in the tensile string and, simply diagonalize the Hamiltonian and obtain the (free string) spectrum and construct the single string Hilbert space and then multi-string Fock space. So, we devise a different procedure: the Schr\"odinger representation quantization. As we will show, the structure of the light-cone Hamiltonian is such that, despite  non-quadratic and non-local features, allows us to get a mileage on constructing the Schr\"odinger wavefunctional and discuss the spectrum.

\subsection{Schr\"odinger wavefunctional and space of physical states } \label{sec:SchrodingerRepresentation}

As the name suggests in our quantization procedure we associate a wavefunctional to our system, which we will denote by\footnote{Note that $x^+$ in the wavefunctional is not a part of a solution parameters and denotes the light-cone time. } 
\begin{equation}
\Psi = \Psi\!\left[ x_0^I, \{X_n^I\}_{n\neq0};\ \bar{p}^+, \bar{x}^1, \bar{p}^1; x^+ \right].
\label{eq:WavefunctionalDef}
\end{equation}
The physical wavefunctional should satisfy Schr\"odinger equation and the null-string level matching condition:
\begin{equation}
H_{\text{LC}} \Psi = i\ \partial_{x^+} \Psi,\qquad G_0 \Psi=0
\label{Schrod-Eq-x+}
\end{equation}
In the $X$ representation,
\begin{equation}
p^I = -i \frac{\partial}{\partial x_0^I}, \qquad
P_n^I = -i \frac{\partial}{\partial X_{-n}^I}.
\label{eq:SchrodingerDifferentialOperators}
\end{equation}
With the above $H_{\text{LC}}$ in the $X$-basis takes the form,
\begin{equation}\label{HLC-X-basis}
H_{\text{LC}} = \frac{1}{2\bar{p}^+} \left[ (\bar{p}^1)^2 - \frac{\partial^2}{\partial x_0^I\partial x_0^I} + \boldsymbol{{\cal M}}^2\right]
\end{equation}
\begin{equation}\label{M2-quantum}
\boldsymbol{{\cal M}^2}= -{\cal G}^{IJ}\ \sum_{n\neq 0} \frac{\delta^2}{\delta X_{-n}^I\delta X_{n}^J}+ \frac{2 x_0^I}{(\bar{x}^1)^2}  \sum_{m,n\neq0} \frac{m}{n} X^J_{n+m} \frac{\delta^2}{\delta X_{m}^I\delta X_{n}^J}+ \frac{1}{(\bar{x}^1)^2} \boldsymbol{\PhiPhi}_\gamma
\end{equation}
with 
\begin{equation}\label{PhiPhi-quantum}
\boldsymbol{\PhiPhi}_\gamma=- \sum_{n,r,s\neq0} \left(\frac{rs}{n^2}\right)\ X^I_{n+r} X^J_{-n+s} \frac{\delta^2}{\delta X_{r}^I\delta X_{s}^J} + \frac{\pi^2}{3}\gamma \sum_{n\neq 0} n^2 X^I_{n} \frac{\delta}{\delta X_{n}^I}
\end{equation}
where we used the fact that $\sum_{n\neq 0} n^{-2}=\frac{\pi^2}{3}$ and $0\leq \gamma \leq 1$ specifies the ordering choice: $\gamma=0$ is ``momentum ordering'' where all the momenta are pushed to the right and $\gamma=1$ is when no ordering is applied, $\boldsymbol{\PhiPhi}\sim X P X P$. As we see, the only term in the Hamiltonian that has an ordering issue is the $\gamma$ term. As we will see, the vacuum state and the physical spectrum and wavefunctionals that are quadratic n $X$, do not depend on $\gamma$.

Assuming that $x^+$ dependence of $\Psi$ is of the form $e^{-iE x^+}$, \eqref{Schrod-Eq-x+} reduces to 
\begin{equation}\label{Schrod-Eq}
H_{\text{LC}} \Psi = E\ \Psi,\qquad G_0 \Psi=0
\end{equation}
(Recall that $[H_{\text{LC}}, G_0]=0$ and can hence be diagonalized simultaneously.) $E$ is the total light-cone energy of the null-string  states described by state/wavefunctional $\Psi$.

\subsection{Vacuum wavefunctional and vacuum energy}\label{sec:VacuumState}

As \eqref{eq:Hamiltonian_Full_Rigorous} explicitly shows, the light-cone Hamiltonian $H_{\text{LC}}$ is positive definite. So, the natural guess is that the vacuum wavefunctional is the one with lowest light-cone energy $E_0=(\bar{p}^1)^2/(2\bar{p}^+)$. Again as \eqref{eq:Hamiltonian_Full_Rigorous} suggests, we define the vacuum state $\Psi_0$ by demanding that it be annihilated by all momentum operators,
\begin{equation}
p^I \Psi_0 = 0, \qquad P_n^I \Psi_0 = 0, \qquad \forall n \neq 0.
\label{eq:VacuumCondition}
\end{equation}
The above then yields $\Phi_n \Psi_0 = 0$. These conditions readily yield $E_0=(\bar{p}^1)^2/(2\bar{p}^+)$, and also $G_0\Psi_0=0$. So, $\Psi_0$ is the lowest lying physical state. 

Using the Schr\"odinger representation \eqref{eq:SchrodingerDifferentialOperators},
the conditions \eqref{eq:VacuumCondition} become
\begin{equation}
\frac{\partial\Psi_0}{\partial x_0^I} = 0, \qquad
\frac{\partial\Psi_0}{\partial X_n^I} = 0, \qquad
\forall n.
\end{equation}
The unique solution is a constant functional,
\begin{equation}
{\Psi_0 = \Psi_0 [\bar{p}^+, \bar{p}^1, \bar{x}^1]=\text{constant}.}
\label{eq:ConstantVacuum}
\end{equation}
The vacuum is therefore the lowest lying  state that is translationally invariant throughout the reduced configuration space. Unlike the tensile-string ground state, it is not localized around a preferred point in configuration space and does not possess a Gaussian profile. This difference reflects the absence of a restoring force in the underlying dynamics. 

Note that the physical wavefunctionals, including the vacuum state, need not be square-integrable in the strict $L^2$ sense over the infinite-dimensional configuration space. To render the functional integration measure $\mathcal{D}X$  well-defined for computing inner products and physical transition amplitudes, we can impose an infrared (IR) finite-volume regulator. This may be achieved by restricting the target-space zero modes and oscillator amplitudes to a large but finite bounding volume $V$, normalizing the vacuum state with respect to this volume, $\Psi_0 \propto V^{-1/2}$. The physical continuum limit $V \to \infty$ is taken strictly at the end of observable calculations.

\subsection{Excited wavefunctionals and the mode-balance condition}\label{sec:PolynomialExcitations}

To work out excited states/wavefunctionals we recall the form of light-cone Hamiltonian in the $X$-basis \eqref{HLC-X-basis}, \eqref{M2-quantum} and \eqref{PhiPhi-quantum}, and the fact that $\boldsymbol{{\cal M}}^2$  generically involves second order derivatives w.r.t $X^I_n$ and there is an ordering $\gamma$-term which involves a first order derivative w.r.t. $X^I_n$.  So,  \eqref{Schrod-Eq} is a second order ODE for zero modes $x_0^I$ and a second order PDE on non-zero mode part of $X^I$. Therefore, one can readily diagonalize the zero mode part by transverse plane-waves:
\begin{equation}
\Psi_{\bar{p}} = \Psi\!\left[\bar{p}_I, \{X_n^I\}_{n\neq0};\ \bar{p}^+, \bar{x}^1, \bar{p}^1; x^+ \right] e^{i\bar{p}_I x_0^I},
\label{eq:WavefunctionalDef}
\end{equation}
for which Schr\"odinger equation \eqref{Schrod-Eq} takes the form,
\begin{equation}\label{M2-mu2}
    \boldsymbol{{\cal M}}^2 \ \Psi_{\bar{p}}=  \mu^2 \Psi_{\bar{p}},\qquad \mu^2:=2\bar{p}^+ E - (\bar{p}^1)^2 - (\bar{p}^I)^2\,.
\end{equation}
$\mu$ can hence be viewed as the target-space mass of the associated null-string state.  The null-string level-matching condition does not involve zero modes and hence $\Psi_{\bar{p}}$ is still subject to the same constraints $G_0 \Psi_{\bar{p}}=0$. 

\paragraph{Monomials wavefunctionals.} One may seek solutions to the eigenvalue equation \eqref{M2-mu2} within polynomials of $X^I_n$. The simplest excitations within this class are monomial functionals built from the Fourier modes,
\begin{equation}
\Psi^K_{\bar{p}} = \prod_{\alpha=1}^{K} X_{n_\alpha}^{I_\alpha}\ \Psi_0(\bar{p}^I; \bar{p}^+, \bar{x}^1, \bar{p}^1).
\label{eq:MonomialState}
\end{equation}
where $\Psi_0$ is an arbitrary function of the argument. Monomial states have the appealing feature that any monomial is an eigenstate of the null-string level-matching operator,
\begin{equation}
G_0 \Psi^K_{\bar{p}} = -i \left( \sum_{\alpha=1}^{K} n_\alpha \right) \Psi^K_{\bar{p}} .
\label{eq:G0Eigenvalue}
\end{equation}
As such, the level-matching constraint can be readily solved for monomial states:
\begin{equation}
{\sum_{\alpha=1}^{K} n_\alpha = 0.}
\label{eq:ModeBalanceCondition}
\end{equation}
The above condition cannot be fulfilled for $K=1$ and hence $K\geq 2$.

As \eqref{M2-quantum} shows, $\boldsymbol{{\cal M}}^2$ has three kind of terms, that  acting on a degree-$K$ state yield a degree-$(K-2)$, a degree-$(K-1)$ and a degree-$K$ state. So, in order a monomial state to be an eigenstate of $\boldsymbol{{\cal M}}^2$, the first two terms in \eqref{M2-quantum} should vanish on the state; the eigenvalues of such states $\mu^2$, cf. \eqref{M2-mu2}, are then specified by the eigenvalues of $\boldsymbol{\PhiPhi}_\gamma$ operator. 

\section{General Quadratic Physical States}\label{sec:quadratic-states}

Within the monomial class, the lowest lying state is the quadratic $K=2$, which upon applying  \eqref{eq:ModeBalanceCondition}, the most general level-matched quadratic state is, 
\begin{equation}
\Psi_2 = \sum_{r\neq0} K^{r}_{IJ} X_{-r}^I X_r ^J\ \Psi_0 .
\label{eq:physicalquadratic}
\end{equation}
The  quadratic wavefunctional is special, because the $X\delta^2/(\delta X\delta X)$ term in \eqref{M2-quantum} vanishes on it, as a result of the level-matching conditions and that \eqref{eq:physicalquadratic} involves $X_{-r}^I X_r ^J$ terms. So, for states in this class to be eigenstate $\boldsymbol{{\cal M}}^2$ they should be zero eigenstates of the $\delta^2/(\delta X\delta X)$ term in \eqref{M2-quantum}, i.e.
\begin{equation}
{\cal G}^{IJ}\ \sum_{n\neq 0} \frac{\delta^2}{\delta X_{-n}^I\delta X_{n}^J}\ \Psi_2 = \sum_{r\neq0} {\cal G}^{IJ}K^{r}_{IJ} \ \Psi_0 =0 \qquad \Longrightarrow \qquad \sum_{r\neq0} {\cal G}^{IJ}K^{r}_{IJ}=0.
\label{eq:PPquadratic}
\end{equation}
The remaining task, which we will take on next, is to diagonalize $\boldsymbol{{\cal M}}^2$ is to determine $K^{r}_{IJ}$ such that $\Psi_2$ is an eigenstate of $\boldsymbol{\PhiPhi}_\gamma$. 

\paragraph{Bilocal representation.} Using
\begin{equation}
X_r^I = \frac{1}{2\pi} \int_0^{2\pi} d\sigma\, e^{-ir\sigma} X^I(\sigma),
\end{equation}
the quadratic state may be written in the bilocal form
\begin{equation}
\Psi_2 = \frac{1}{(2\pi)^2}\ \int_0^{2\pi} d\sigma \int_0^{2\pi} d\tilde{\sigma}\, X^I(\sigma) X^J(\tilde{\sigma}) K_{IJ}(\sigma,\tilde{\sigma}),
\label{eq:BilocalKernel}
\end{equation}
where
\begin{equation}
K_{IJ}(\sigma,\tilde{\sigma}) = \sum_{r\neq0} K^r_{IJ} e^{-ir(\sigma-\tilde{\sigma})}.
\end{equation}
The mode-balancing condition implies that the kernel depends only on the relative coordinate. Introducing
\begin{equation}
\theta \equiv \sigma-\tilde{\sigma} \;\ (\mathrm{mod}\; 2\pi), \qquad 0\leq\theta<2\pi,
\end{equation}
we may write
\begin{equation}
K_{IJ}(\sigma,\tilde{\sigma}) := K_{IJ}(\theta),
\end{equation}
with
\begin{equation}
K_{IJ}(\theta) = \sum_{r\neq0} K^r_{IJ} e^{-ir\theta}.
\label{eq:Btheta}
\end{equation}
Thus $K_{IJ}$ is naturally a function on the relative circle $S^1$ and satisfies
\begin{equation}
\boxed{\ K_{IJ}(\theta+2\pi) = K_{IJ}(\theta) \ }
\label{eq:Bperiodic}
\end{equation}

\subsection{Exchange symmetry and tensor decomposition}

The variables $\sigma$ and $\tilde{\sigma}$ are dummy integration variables. Therefore the wavefunctional must be invariant under
\begin{equation}
(\sigma,I) \longleftrightarrow (\tilde{\sigma},J).
\end{equation}
In terms of the relative coordinate,
\begin{equation}
\theta \longrightarrow 2\pi-\theta,
\end{equation}
and therefore
\begin{equation}
\boxed{\ K_{IJ}(\theta) = K_{JI}(2\pi-\theta)\ }
\label{eq:ExchangeSymmetry}
\end{equation}

The kernel may be decomposed into symmetric-traceless, antisymmetric and trace sectors with respect to metric ${\cal G}_{IJ}$,
\begin{equation}
\begin{split}
K_{IJ} = G_{IJ} + B_{IJ} + \frac1d{\cal G}^{-1}_{IJ}\ \varphi,  \qquad d=D-3. \\ ({\cal G}^{-1})^{IJ}=  \delta^{IJ}- \frac{1}{(\bar{x}^1)^2+ (x_0^K)^2}x_0^Ix_0^J ,\qquad {\cal G}^{IJ} G_{IJ}=0,    
\end{split}\end{equation}
Equation \eqref{eq:ExchangeSymmetry} implies
\begin{equation}
\begin{split}\label{G-B-varphi-AS-S}
G_{IJ}(\theta) &= G_{JI}(\theta) =G_{IJ}(2\pi-\theta), \\
B_{IJ}(\theta) &= -B_{JI}(\theta)= -B_{IJ}(2\pi-\theta), \\
\varphi(\theta) &= \varphi(2\pi-\theta).
\end{split}
\end{equation}
and \eqref{eq:PPquadratic} implies,
\begin{equation}\label{varphi-0}
    \varphi(0)=\varphi(2\pi)=0\,.
\end{equation}
Thus the symmetric-traceless and trace sectors are even under reflection about $\theta=\pi$, while the antisymmetric sector is odd.

Next, we should explore the quadratic states as eigenstates of the Hamiltonian, i.e. eigenstates of $\boldsymbol{\PhiPhi}_\gamma$. 

\subsection{Quadratic eigen-wavefunctionals}

Given $\boldsymbol{\PhiPhi}_\gamma$ in \eqref{PhiPhi-quantum}, we have
\begin{equation}
\boldsymbol{\PhiPhi}_\gamma\ \Psi_2 =  \frac{2}{(2\pi)^2} \left[\int_0^{2\pi} d\sigma \int_0^{2\pi} d\tilde{\sigma}\, X^I(\sigma) X^J(\tilde{\sigma}) \beta_\gamma(\theta) \partial_\theta^2 K_{IJ}(\theta)\right]\ \Psi_0,
\end{equation}
where
\begin{equation}
\beta_\gamma(\theta) = \sum_{n\neq0} \frac{e^{in\theta}}{n^2}-\frac{\pi^2}{3}\gamma = 2\sum_{n=1}^{\infty} \frac{\cos(n\theta)}{n^2}-\frac{\pi^2}{3}\gamma, \qquad 0\leq \gamma \leq 1,
\label{eq:betaFourier}
\end{equation}
where $\gamma$ is the ordering factor.  On the fundamental domain $0\leq\theta\leq2\pi$ this series sums to
\begin{equation}
\beta_\gamma(\theta) = \frac{\pi^2}{3}(1-\gamma) - \pi\theta + \frac{\theta^2}{2}=\frac12\big[(\theta-\pi)^2-\frac{\pi^2}{3}(1+2\gamma)\big].
\label{eq:betaPolynomial}
\end{equation}

Requiring $\Psi_2$ to be an eigenstate gives
\begin{equation}\label{eq:BbetaEq}
\boxed{\  \beta_\gamma(\theta) \partial_\theta^2 K_{IJ}(\theta)=\frac12\lambda (\lambda+1)\ K_{IJ}(\theta)  \ }
\end{equation}
where for later convenience we have parametrized the eigenvalue in terms of $\lambda$, which is related to the mass eigenvalue $\mu^2$, cf. \eqref{M2-mu2},  as 
\begin{equation}
   \bar{x}_1^2 \mu^2= \lambda (\lambda+1)
\end{equation}
The above is  linear in $K_{IJ}$ and therefore, each of the $G_{IJ}, B_{IJ}$ and $\varphi$ sectors should satisfy this equation. Moreover, they are subject to  \eqref{eq:Bperiodic} and \eqref{G-B-varphi-AS-S}. We explore these conditions next. 

\paragraph{Solutions to the ODE \eqref{eq:BbetaEq}.} 
For later convenience, it is helpful to introduce a coordinate centered on the reflection point:
\begin{equation}
t = \frac1\alpha (\theta-\pi),\qquad \alpha^2=\frac{\pi^2}{3}(1+2\gamma), \qquad 
\qquad -t_\gamma\leq t\leq t_\gamma ,\qquad t_\gamma:=\sqrt{\frac{3}{1+2\gamma}}.
\label{eq:xCoordinate}
\end{equation}
We note that $1\leq t_\gamma\leq \sqrt{3}$. 
In this coordinate \eqref{eq:BbetaEq} is written as,
\begin{equation}
(1-t^2)K''(t)+\lambda(\lambda+1) K(t)=0\,,
\label{eq:traceODEt}
\end{equation}
and the periodicity and reflection  conditions take the form, 
\begin{equation}\label{periodicity-cond}
K_{IJ}(t-t_\gamma) = K_{IJ}(t+t_\gamma),\qquad  \forall t
\end{equation}
which, once one restricts oneself to $t\in [-t_\gamma, +t_\gamma]$ region, implies that the function $K_{IJ}(t)$ should be smooth at $t=\pm t_\gamma$ points and  
\begin{equation}\label{smoothness-at-sqrt3}
K^{(n)}_{IJ}(-t_\gamma ) = K_{IJ}^{(n)}(t_\gamma),\qquad  \forall n
\end{equation}
where $K^{(n)}_{IJ}$ denotes $n^{\text{th}}$ derivative of the function. Finally, \eqref{G-B-varphi-AS-S} and \eqref{varphi-0} take the form,
\begin{align}\label{odd-even-B}
G_{IJ}(-t) = G_{IJ}(t),  \qquad  B_{IJ}(-t) = -B_{IJ}(t), \qquad \varphi(-t) = \varphi (t),\qquad \varphi (t_\gamma)=0.
\end{align}

Upon a simple transformation, 
\begin{equation}
K(t) = \sqrt{1-t^2} f(t),
\label{eq:BtoF}
\end{equation}
\eqref{eq:traceODEt} leads to the standard associated Legendre equation \cite{Gradshteyn-Ryzhik}:
\begin{equation}
(1-t^2)f''(t) - 2t f'(t) + \left[ \lambda(\lambda+1) - \frac{1}{1-t^2} \right] f(t) = 0.
\label{eq:assocLegendreDerived}
\end{equation}
and its solutions are associated Legendre functions $P^m_\lambda(t), Q^m_\lambda(t)$ with $m=\pm1$.

\paragraph{Special $\lambda=0$ case. } For $\lambda=0$, \eqref{eq:traceODEt} reduces to
\begin{equation}
(1-t^2)K''(t)=0,
\end{equation}
with general solution
\begin{equation}
K''(t)=0 \qquad \Longrightarrow \qquad K(t)=A+Ct.
\end{equation}
It is easy to see that the only solution that admits continuity and periodicity conditions of the previous subsection are $K(t)=0$. That is, the vacuum state is the only state with $\lambda=0$.  So, we only focus on $\lambda\neq 0, -1$ solutions.

\paragraph{General solutions.}

On any interval not containing the regular singular points $t=\pm1$,\footnote{It is interesting to note that the  $t=\pm t_\gamma$ correspond to $\theta=0,2\pi$, directly connected to having a closed string, whereas $t=\pm 1$ arise from the sum in \eqref{eq:betaFourier} and  $\sqrt{\frac{3}{\pi}\sum_{n=1} {n^{-2}}}$ } the solution space is a general linear combination of the two linearly independent associated Legendre functions, 
\begin{equation}
K(t) = \sqrt{1-t^2} \left[ A P_\lambda^1(t) + C Q_\lambda^1(t) \right],
\label{eq:BGeneralLegendre}
\end{equation}
Note that, $P_\lambda^{-1}(t), Q_\lambda^{-1}(t)$ are a linear combination of $P_\lambda^1(t), Q_\lambda^1(t)$. Equation \eqref{eq:BGeneralLegendre} should be viewed as a \textit{local representation}. Since the physical domain contains the regular singular points $t=\pm1$, the admissibility of a given solution depends on its behavior near those points and on the matching conditions imposed there. To formally analyze the solution in the full $t\in [-t_\gamma, t_\gamma]$ range, we write the solution as
\begin{equation}\label{B-full-range}
    K(t)=\left\{\begin{array}{cc} \sqrt{1-t^2} \left[ A_- P_\lambda^1(t) + C_- Q_\lambda^1(t) \right] &\qquad -t_\gamma\leq t \leq -1 \\  & \\ \sqrt{1-t^2} \left[ A_0 P_\lambda^1(t) + C_0 Q_\lambda^1(t) \right] &\qquad -1\leq t \leq +1 \\ & \\ \sqrt{1-t^2} \left[ A_+ P_\lambda^1(t) + C_+ Q_\lambda^1(t) \right] &\qquad +1\leq t \leq +t_\gamma
    \end{array} \right.
\end{equation}
The same solution is required to be valid in the whole range of $t$ (beyond $t\in [-t_\gamma, t_\gamma]$ region) by extending the above using \eqref{periodicity-cond}, subject to smoothness conditions at $t=\pm t_\gamma$ points \eqref{smoothness-at-sqrt3}.

\paragraph{Admissible solutions, discreteness of the spectrum.}
\label{sec:continuity}

We consider the three different cases of $K=K(t)$ and impose the relevant periodicity and continuity requirements, \eqref{periodicity-cond} and \eqref{odd-even-B}. We present the detail of the analysis for one case (antisymmetric case) and skip the details for the other two, which are quite similar. 

\paragraph{\vspace*{3mm} (1) Antisymmetric Sector.} In the middle region $-1\leq t \leq 1$, we can write the odd solution as
\begin{equation}
    K_{odd}(t)=\sqrt{1-t^2}\big(A_0\ P^1_\lambda(t)+C_0\ Q^1_\lambda (t)\big).
\end{equation}
We have \cite{Gradshteyn-Ryzhik}
\begin{equation}
\begin{split}
    P^1_\lambda(-t)&=-\cos(\lambda\pi)\ P^1_\lambda(t)+\frac{2}{\pi}\sin( \lambda\pi)\ Q^1_\lambda(t),\\
    Q^1_\lambda(-t)&=\cos(\lambda\pi)\ Q^1_\lambda(t)+\frac{\pi}{2}\sin(\lambda\pi)\ P^1_\lambda(t).
\end{split}
\end{equation}
Requiring $K(-t)=-K(t)$, we find
\begin{equation}
    {A_0}\sin(\frac{\lambda\pi}{2})=-\frac{\pi}{2} C_0\cos(\frac{\lambda\pi}{2}).
\end{equation}
So, 
\begin{equation}\begin{split}
    C_0&=0,\  A_0\neq 0 \qquad\qquad  \lambda=2k\\
 A_0&=0,\  C_0\neq 0 \qquad\qquad \lambda=2k+1\\    
 C_0&=-\frac{2}{\pi}A_0 \tan(\frac{\lambda\pi}{2}) \qquad \lambda\notin \mathbb{Z}
\end{split}
\end{equation}

In the left and right regions, we can write 
\begin{equation}
\begin{split}
    K_+(t)&=\sqrt{1-t^2}\big(A_+\ P^1_\lambda(t)+C_+ \ Q^1_\lambda(t)\big), \qquad 1\leq t\leq t_\gamma\\
    K_-(t)&=\sqrt{1-t^2}\big(A_-\ P^1_\lambda(t)+C_-\ Q^1_\lambda(t)\big),\qquad -t_\gamma\leq t\leq -1    
\end{split}
\end{equation}
Requiring $K$ to be odd we find
\begin{equation}\label{lrcons}
\begin{split}
 A_+=   A_- \cos(\lambda\pi)-\frac{\pi}{2}C_-\sin(\lambda\pi), \qquad C_+=-\frac{2}{\pi}A_- \sin(\lambda\pi)-C_-\cos(\lambda\pi).
\end{split}\end{equation}
Requiring $K(t_\gamma)=K(-t_\gamma)=0$, yields
\begin{equation}\label{C+A+C-A_}
    C_+=-\frac{P^1_\lambda(t_\gamma)}{Q^1_\lambda(t_\gamma)}\ A_+,\qquad  C_-=-\frac{P^1_\lambda(-t_\gamma)}{Q^1_\lambda(-t_\gamma)}\ A_-\,.
\end{equation}
Since $K$ is odd, there is no extra condition from derivative of $K$ at $\pm t_\gamma$. With the above we have 4 relations among 6 coefficients and hence we have two independent coefficients. Note that given \eqref{lrcons}, only one of the relations in \eqref{C+A+C-A_} is independent. 
Since $K$ is defined up to an overall factor, therefore, there is only one independent coefficient.

To study continuity at $t=\pm 1$, we note that
\begin{equation}
\begin{split}    
\sqrt{1-t^2}Q^1_\lambda(t)|_{t\rightarrow 1}=-1, &\qquad  \sqrt{1-t^2}Q^1_\lambda(t)|_{t\rightarrow -1}=-\cos(\lambda\pi),\\
    \sqrt{1-t^2}P^1_\lambda(t)|_{t\rightarrow 1}=0, &\qquad  \sqrt{1-t^2}P^1_\lambda(t)|_{t\rightarrow -1}=-\frac{2}{\pi}\sin(\lambda\pi).
\end{split}\end{equation}
From these, continuity at $t=\pm 1$ yields
\begin{equation}
    C_+=C_0, 
\end{equation}
With the above $K$ has been determined up to an overall factor, with no restriction on $\lambda$.

So far, we have only imposed the continuity conditions within $t\in (-t_\gamma, t_\gamma)$. However, what we actually have is \eqref{smoothness-at-sqrt3} which is stronger. 
Recalling properties of associated Legendre functions we see that this is only possible, iff $C_+=0=A_+$ and hence $C_-=0=A_-$. This yields $C_0=0$. Therefore, to have a non-trivial solution, we must have $A_0\neq 0$, which is possible iff $\lambda$ is an even integer. All in all, the admissible solution in the antisymmetric part is, 
\begin{equation}\label{B-AS-full-range}
    B_{IJ}(t)=\left\{\begin{array}{cc} 0 &\qquad -t_\gamma\leq t \leq -1  \\ B^0_{IJ}\ \sqrt{1-t^2} \ P_{2k}^1(t)    &\qquad -1\leq t \leq +1 \\ 0  &\qquad +1\leq t \leq +t_\gamma 
    \end{array} \right.
\end{equation}
where $B_{IJ}^0$ is a given constant anti-symmetric tensor with the discrete spectrum,
\begin{equation}\label{mass-AS-case}
    (\bar{x}_1)^2 \mu^2= k(2k+1),\qquad k=1,2,\cdots  
\end{equation}
with $\mu$ given in \eqref{M2-mu2}. As we see  $\bar{x}_1$ acts as ``effective'' null-string length scale. 

\paragraph{\vspace*{3mm} (2) Trace Sector.}
In this case, $K(t)=K(-t)$ and
\begin{equation}
\begin{split}	
\varphi(\pm t_\gamma)=0,\qquad 
\varphi'(\pm t_\gamma)=0.
\end{split}
\end{equation}
Recall that we are dealing with a second order ODE with $t=\pm t_\gamma $ as regular points, thus $\varphi (\pm t_\gamma)=0, \varphi'(\pm t_\gamma)=0$, imply that 
\begin{equation}
\varphi(t)=0 \ \ \textrm{ for}\qquad 1\leq t\leq t_\gamma \ \text{ or }\ -t_\gamma \leq t\leq -1.
\end{equation}
Therefore, if the kernel is assumed to be continuous, one must have
\begin{equation}
\varphi(1)=\varphi(-1)=0.
\label{eq:TraceSingularConditions}
\end{equation}
Since $t=\pm1$ are regular singular points of the differential equation, \eqref{eq:TraceSingularConditions} admits a nontrivial solution on the interior interval $-1<t<1$. We need to impose \eqref{eq:TraceSingularConditions} on \eqref{B-full-range}. Since $\varphi(\pm1)=0$ and noting that $Q_\lambda^1$-branch does not satisfy the endpoint condition, then $C_0=0$ and we remain with, 
\begin{equation}
\varphi(t) = \sqrt{1-t^2}\ P_\lambda^1(t).
\label{eq:TracePBranch}
\end{equation}
Requiring $P_\lambda^1(t)=P_\lambda^1(-t)$ implies that $\sin(\lambda\pi)=0, \cos(\lambda\pi)=-1$, therefore, $\lambda$ must be an odd integer,  $\lambda=2k-1, k=1,2, \cdots $ and hence 
\begin{equation}\label{B-Trace-full-range}
    \varphi(t)=\left\{\begin{array}{cc} 0 &\qquad -t_\gamma\leq t \leq -1  \\ \sqrt{1-t^2}P^1_{2k-1}(t)    &\qquad -1\leq t \leq +1 \\ 0  &\qquad +1\leq t \leq +t_\gamma 
    \end{array} \right.
\end{equation}
For this case, 
\begin{equation}\label{mass-Trace-case}
    (\bar{x}_1)^2 \mu^2= k(2k-1),\qquad k=1,2,\cdots  
\end{equation}

\paragraph{\vspace*{3mm} (3) Symmetric Traceless Sector.} As similar analysis for this case again yields 
\begin{equation}\label{B-Sym-full-range}
    K(t)=\left\{\begin{array}{cc} 0 &\qquad -t_\gamma\leq t \leq -1  \\ \sqrt{1-t^2}\big(A_0 P^1_\lambda(t)+C_0 Q^1_\lambda (t)\big)    &\qquad -1\leq t \leq +1 \\ 0  &\qquad +1\leq t \leq +t_\gamma 
    \end{array} \right.
\end{equation}
where $\sqrt{1-t^2}\big(A_0 P^1_\lambda(t)+C_0 Q^1_\lambda (t)\big)$ should be an even function. This requirement yields
\begin{eqnarray}
    A_0 \cos(\frac{\lambda\pi}{2})=\frac{\pi}{2} C_0 \sin(\frac{\lambda\pi}{2})
\end{eqnarray}
Continuity at $t=1$, implies $C_0=0$. Therefore, to have $A_0\neq 0$ we should have $\lambda$ should an odd integer, $\lambda=2k-1, k=1,2, \cdots $ and hence 
\begin{equation}\label{GIJ-Sym-full-range}
    G_{IJ}(t)=\left\{\begin{array}{cc} 0 &\qquad -t_\gamma\leq t \leq -1  \\ G_{IJ}^0\ \sqrt{1-t^2}P^1_{2k-1}(t)    &\qquad -1\leq t \leq +1 \\ 0  &\qquad +1\leq t \leq +t_\gamma 
    \end{array} \right.
\end{equation}
where $G_{IJ}^0$ is a constant symmetric tensor with ${\cal G}^{IJ}G_{IJ}^0=0$. For this case, 
\begin{equation}\label{mass-Sym-Traceless-case}
    (\bar{x}_1)^2 \mu^2= k(2k-1),\qquad k=1,2,\cdots  
\end{equation}

\paragraph{Summary of the section.} Symmetric and antisymmetric parts of $K_{IJ}$ are respectively given by odd and even  associated Legendre polynomial $P_\ell^1(t)$ for $-1\leq t\leq 1$ and vanish in the rest of $[-t_\gamma,+t_\gamma]$ region, where $t^2_\gamma=\frac{3}{1+2\gamma}, 0\leq \gamma\leq 1$ is the only ordering dependent quantity in our analysis. For all these cases, therefore, the spectrum is discrete $2(\bar{x}_1)^2 \mu^2=\ell(\ell+1)$. We crucially note that the physical spectrum and also the non-zero part of the quadratic wavefunctional are independent of $\gamma$.

\section{Discussion and Outlook}\label{sec:Outlook}

We have provided a complete and self-contained canonical quantization of the closed null-string on flat $D$ dimensional target-space  in the ``null-string light-cone gauge'' (NSLCG),  carefully incorporating the recently identified local Carroll-Weyl scaling symmetry \cite{Sheikh-Jabbari:2026cnj, Sheikh-Jabbari:2026vqh}. The light-cone Hamiltonian is explicitly positive definite and is free of ordering ambiguity (cf. discussions below \eqref{XP-dagger}), while is highly non-local. The non-locality arises as a result of imposing the $\texttt{C}_3$ constraint which arises from the ``overlooked gauge symmetry'' of the null strings \cite{Sheikh-Jabbari:2026cnj}. We performed the canonical quantization in the Schr\"odinger representation and discussed that \eqref{Schrod-Eq} provides the equations  any physical wavefunctional should satisfy. Finding eigen wavefunctionals amounts to solving a second order Schr\"odinger PDEs. We  uniquely specified the lowest energy states, the vacuum state. 

We then discussed a general class of monomial wavefunctions of the form \eqref{eq:MonomialState}. The reason to consider monomial states is the fact that one can easily impose the null-string level matching constraint. We thoroughly analyzed the quadratic wavefunctionals and found the spectrum of the null-strings in this class. We found a perhaps unexpected result that these states have a discrete spectrum; unexpected because null-strings may be viewed as a congruence of \textit{null geodesics} in the flat $D$ dimensional target-space, subject to the three constraints. This apparent puzzle can be understood once we recall that the non-zero ``mass'' for the null string is coming from the $\boldsymbol{\PhiPhi}_\gamma$ term in the light-cone Hamiltonian \eqref{eq:Hamiltonian_Full_Rigorous} and involves highly nonlocal terms. The $\bar{x}^1$, which is the solution parameter associated with the value $X^{1}(\sigma)$, appears as effective string tension in the quadratic wavefunctional sector.  

\paragraph{Further comments on the quadratic wavefunctionals.} The tower of massive states we found in the quadratic sector, from the target-space viewpoint, are associated with symmetric-traceless, anti-symmetric two tensors in the $D-3$-dimensional transverse space, as well as a scalar. This parallels the massless NSNS sector of tensile strings \cite{Polchinski:1998rq},  with two important differences that here we have a tower of massive strings and that these are carrying  $D-3$ dimensional polarizations (in contrast to $D-2$ of the tensile strings). Moreover, it is illuminating to recall  the bilocal character of the quadratic wavefunctional and that the kernel function $K_{IJ}(\sigma-\tilde\sigma)$ is only non-zero for $-\pi \sqrt{\frac{1+2\gamma}{3}} < \sigma-\tilde\sigma <\pi \sqrt{\frac{1+2\gamma}{3}}$. Therefore, the wavefunctional involves non-local product of string fields $X^I(\sigma)$ and $X^J(\tilde\sigma)$. The nonlocality of the Hmailtonian and the wavefunctional which is a result of imposing the $\texttt{C}_3$ constraint, is how one restores an effective string tension and hence interactions on a Carrollian worldsheet. Note that on a Carrollian worldsheet all points on a constant time slice are causally disconnected and the only way to have (covariant) interaction terms is through introducing nonlocality in $\sigma$.

Within our framework we obtained two basic results for closed null-strings on flat $D$ dimensional target space:
(1) lack of critical dimension and (2) discrete spectrum. Finding the discrete spectrum calls into question the lore that null-strings and higher-spin gauge theories should somehow be connected to each other see e.g. \cite{Bonelli:2003kh,  Sundborg:2000wp, Fotopoulos:2010ay}. It is instructive to verify these results in more general contexts, considering non-flat backgrounds, open strings and going beyond the quadratic wavefunctionals. In particular, one can ask what are the consistent null-string backgrounds.

\paragraph{Other quantization schemes?} The reduced light-cone Hamiltonian lacks the universal quadratic potential characteristic of the tensile string. Thus, the standard oscillator Fock-space construction is unavailable. We have demonstrated that the Schr\"odinger representation provides the natural and robust alternative for the tensionless limit, where quantum states are realized as wavefunctionals on the reduced configuration space. Within this framework, the vacuum state manifests as a translationally invariant constant functional with identically vanishing zero-point energy, reflecting a complete absence of oscillator intercepts. However, one may try quantizing the theory covariantly, without fixing the NSLCG. We then need to extend the standard $bc$-ghosts \cite{Polchinski:1998rq} to $bcs$-ghost system \cite{Sheikh-Jabbari:2026tpf, Duary:2026rlo}. It is an important consistency check to verify that the main results we discussed here are independent of the quantization scheme.

\begin{acknowledgments}
We thank Shing-Tung Yau and Alireza Akbari for discussions. MMShJ acknowledges Iranian National Science Foundation (INSF) research chair grant No.40451653. The work of HY is supported in part by Beijing Natural Science Foundation under Grant No. IS23013.
\end{acknowledgments}

\appendix

\appendix
\section{Closed Bosonic Strings in the Schr\"odinger Representation}
\label{app:Tensile}

In this appendix, we review the light-cone quantization of the standard tensile closed bosonic string in the Schr\"odinger representation. The purpose of this discussion is to explicitly exhibit the canonical phase-space geometry and the resulting Gaussian wavefunctional. This provides a rigorous point of comparison for the Carrollian null-string considered in the main text, where the standard Fock space construction is no longer naturally applicable.

We do not review the Lorentz-algebra anomaly or critical-dimension analysis ($D=26$) here; this appendix is strictly meant to compare the Schr\"odinger wavefunctional structure of the tensile and null strings. Furthermore, in the following we display the formal Schr\"odinger vacuum and suppress the usual normal-ordering constant (zero-point energy).

\subsection{Canonical phase space and mode expansion}

We begin with a closed bosonic string propagating in flat Minkowski spacetime, with worldsheet coordinates $(\tau, \sigma)$, where $\sigma \in [0, 2\pi]$. We use light-cone coordinates with the target-space metric convention $ds^2 = -2 dX^+ dX^- + \delta_{IJ} dX^I dX^J$. 

Imposing the light-cone gauge $X^+(\tau,\sigma) = x^+ + \alpha' p^+ \tau$, the Virasoro constraints determine the nonzero modes of the longitudinal coordinate $X^-$, while the zero-mode relation determines $p^-$ in terms of the transverse Hamiltonian. 

To transition to the Schr\"odinger picture, we evaluate the transverse embedding coordinates $X^I(\sigma)$ (for $I = 1, \dots, D-2$) and their conjugate momentum densities $P^I(\sigma)$ at a fixed worldsheet time, $\tau = 0$. Because the string is closed, we enforce the periodicity condition $X^I(\sigma+2\pi) = X^I(\sigma)$. We decompose the fields into their spatial Fourier modes:
\begin{equation}
X^I(\sigma) = \frac{1}{\sqrt{2\pi}} \sum_{n\in\mathbb{Z}} X_n^I e^{i n \sigma}, \qquad P^I(\sigma) = \frac{1}{\sqrt{2\pi}} \sum_{n\in\mathbb{Z}} P_n^I e^{i n \sigma}.
\end{equation}
The reality of the classical string coordinates imposes $(X_n^I)^\ast = X_{-n}^I$ and $(P_n^I)^\ast = P_{-n}^I$. The equal-time canonical commutation relations $[X^I(\sigma), P^J(\sigma')] = i \delta^{IJ} \delta(\sigma - \sigma')$ translate directly to the discrete mode algebra:
\begin{equation}
[X_n^I, P_m^J] = i \delta^{IJ} \delta_{n+m, 0}.
\end{equation}

With this normalization convention, the zero-modes $X_0^I$ and $P_0^I$ are related to the standard center-of-mass coordinate $x^I$ and total transverse momentum $p^I$ by
\begin{equation}
x^I = \frac{1}{2\pi} \int_0^{2\pi} d\sigma \, X^I(\sigma) = \frac{X_0^I}{\sqrt{2\pi}}, \qquad p^I = \int_0^{2\pi} d\sigma \, P^I(\sigma) = \sqrt{2\pi}P_0^I.
\end{equation}

The canonical phase-space geometry is defined by the symplectic two-form:
\begin{equation}
\Omega = \int_0^{2\pi} d\sigma \, \delta X^I(\sigma) \wedge \delta P^I(\sigma) = \sum_{n \in \mathbb{Z}} \delta X_n^I \wedge \delta P_{-n}^I.
\end{equation}
In the second expression, we use the complexified Fourier-mode notation; the real phase space is recovered by imposing the reality conditions $(X_{-n}^I)^\ast = X_n^I$ and $(P_{-n}^I)^\ast = P_n^I$.

\subsection{Level matching and the light-cone Hamiltonian}

For a closed string, after light-cone gauge fixing, the Virasoro constraints determine $X^-$ only if the periodicity condition for $X^-$ is satisfied. In phase-space form, this requires
\begin{equation}
\mathcal{P}_\sigma = \int_0^{2\pi} d\sigma \, P^I(\sigma)\partial_\sigma X^I(\sigma) \approx 0.
\end{equation}
In terms of the Fourier modes, this becomes $i\sum_{n\in\mathbb{Z}} n P_{-n}^I X_n^I = 0$. At the quantum level, this expression is understood with the usual normal ordering, which in the oscillator language translates to the familiar condition $N - \widetilde{N} = 0$. This gives the global residual condition usually known as level matching; it is the global remnant of the spatial reparametrization constraint. This residual closed-string condition is structurally important and will be compared with the residual/global constraints appearing in the null-string framework discussed in the main text.

The dynamics are governed by the transverse worldsheet Hamiltonian:
\begin{equation}
H_{\text{lc}} = \frac{1}{2} \int_0^{2\pi} d\sigma \left( 2\pi\alpha' P^I(\sigma) P^I(\sigma) + \frac{1}{2\pi\alpha'} \partial_\sigma X^I(\sigma) \partial_\sigma X^I(\sigma) \right).
\end{equation}
With these conventions, the classical relation is $H_{\text{lc}} = \alpha' p^+ p^-$. Quantum mechanically, this relation is shifted by the usual normal-ordering constant. Substituting the Fourier expansions into the Hamiltonian yields:
\begin{equation}
H_{\text{lc}} = \frac{\alpha'}{2}(p^I)^2 + \sum_{n>0} \left( 2\pi\alpha' P_n^I P_{-n}^I + \frac{n^2}{2\pi\alpha'} X_n^I X_{-n}^I \right).
\end{equation}
Note that the zero-mode term evaluates exactly to $\frac{\alpha'}{2}(p^I)^2$, and we have rewritten the sum over strictly positive integers $n>0$ to make the degree-of-freedom counting transparent.

\subsection{Schr\"odinger representation and the vacuum}

To quantize the theory in the Schr\"odinger representation, we define wavefunctionals $\Psi = \Psi[x^I, \{X_n^I\}_{n \neq 0}]$, with the complex modes understood as shorthand for the corresponding real sine and cosine variables. In this complex Fourier notation, $X_n^I$ and $X_{-n}^I$ are treated as complex-conjugate configuration variables; equivalently, one may pass to real sine and cosine modes. The momentum operators are defined as:
\begin{equation}
P_n^I = -i \frac{\partial}{\partial X_{-n}^I} \quad (n \neq 0), \qquad p^I = -i \frac{\partial}{\partial x^I}.
\end{equation}
The Hamiltonian becomes an infinite sum of decoupled harmonic-oscillator differential operators:
\begin{equation}
H_{\text{lc}} = -\frac{\alpha'}{2} \frac{\partial^2}{\partial x^I \partial x^I} + \sum_{n>0} \left( -2\pi\alpha' \frac{\partial^2}{\partial X_n^I \partial X_{-n}^I} + \frac{n^2}{2\pi\alpha'} X_n^I X_{-n}^I \right).
\label{eq:SchrodingerHamiltonian}
\end{equation}

For each $n>0$, the pair $(X_n^I, P_n^I)$, together with its complex conjugate, is equivalent to two real harmonic oscillators. In the usual closed-string notation, linear combinations of these phase-space variables are repackaged as the independent left- and right-moving oscillator modes $\alpha_n^I$ and $\widetilde{\alpha}_n^I$. 

The transverse oscillator vacuum at fixed momentum, denoted as $|0, p\rangle$, is the Gaussian eigen functional of the normal-ordered oscillator Hamiltonian. Equivalently, it is the exact solution annihilated by all oscillator lowering operators. Writing these first-order conditions directly as differential equations on the wavefunctional, we require:
\begin{equation}
\left( \frac{\partial}{\partial X_{-n}^I} + \frac{n}{2\pi\alpha'} X_n^I \right)\Psi_0 = 0, \qquad \left( \frac{\partial}{\partial X_n^I} + \frac{n}{2\pi\alpha'} X_{-n}^I \right)\Psi_0 = 0, \qquad \forall n>0.
\end{equation}
The differential operator in \eqref{eq:SchrodingerHamiltonian} has the usual formal zero-point energy; normal ordering subtracts this constant and leaves the Gaussian wavefunctional unchanged. The formal solution of this infinite set of equations gives the transverse oscillator vacuum wavefunctional:
\begin{equation}
\Psi_0[x^I, X_n^I] = \mathcal{N} e^{i p^I x^I} \exp\left( - \frac{1}{2\pi\alpha'} \sum_{n>0} n \, X_n^I X_{-n}^I \right).
\end{equation}
Here, the center-of-mass factor $e^{ip^I x^I}$ is understood in the usual plane-wave (delta-function) normalization, and the normalization constant $\mathcal{N}$ is formal, requiring the usual regularization of the infinite product over oscillator modes.

This wavefunctional reveals that the transverse embedding coordinates of the tensile string undergo Gaussian fluctuations, where the width of the $n$-th oscillator scales as $(\alpha'/n)^{1/2}$. Formally, the Gaussian width diverges as $\alpha' \to \infty$, reflecting the disappearance of the tensile restoring force. The actual null-string limit is singular at the Hamiltonian level (as the $P^2$ term diverges if $P$ is held fixed). Thus, while the tensile string Schr\"odinger vacuum is governed by a Gaussian oscillator measure, the null string requires a Schr\"odinger-type treatment in which the physical state space is determined by the Carrollian constraint system derived in the main text.


\begin{thebibliography}{99}

\bibitem{Sheikh-Jabbari:2026cnj}
M.~M. Sheikh-Jabbari and H.~Yavartanoo, ``{On the Consistency of Null Strings Literature: The Tale of an Overlooked Symmetry},'' \href{http://www.arXiv.org/abs/2605.12414}{{\tt 2605.12414}}.

\bibitem{Sheikh-Jabbari:2026vqh}
M.~M. Sheikh-Jabbari and H.~Yavartanoo, ``{Null Strings Gauged and Reloaded, I: Null Strings Have Carroll-Weyl Gauge Symmetry},'' \href{http://www.arXiv.org/abs/2605.25817}{{\tt 2605.25817}}.

\bibitem{Bagchi:2026wcu}
A.~Bagchi, A.~Banerjee, R.~Chatterjee, and P.~Pandit, ``{The Tensionless Lives of Null Strings},'' \href{http://www.arXiv.org/abs/2601.20959}{{\tt 2601.20959}}.

\bibitem{Schild:1976vq}
A.~Schild, ``{Classical Null Strings},'' {\em Phys. Rev. D} {\bf 16} (1977) 1722.

\bibitem{Lindstrom:1990qb}
U.~Lindstrom, B.~Sundborg, and G.~Theodoridis, ``{The Zero tension limit of the superstring},'' {\em Phys. Lett. B} {\bf 253} (1991) 319--323.

\bibitem{Isberg:1992ia}
J.~Isberg, U.~Lindstrom, and B.~Sundborg, ``{Space-time symmetries of quantized tensionless strings},'' {\em Phys. Lett. B} {\bf 293} (1992) 321--326, \href{http://www.arXiv.org/abs/hep-th/9207005}{{\tt hep-th/9207005}}.

\bibitem{Isberg:1993av}
J.~Isberg, U.~Lindstrom, B.~Sundborg, and G.~Theodoridis, ``{Classical and quantized tensionless strings},'' {\em Nucl. Phys. B} {\bf 411} (1994) 122--156, \href{http://www.arXiv.org/abs/hep-th/9307108}{{\tt hep-th/9307108}}.

\bibitem{Gustafsson:1994kr}
H.~Gustafsson, U.~Lindstrom, P.~Saltsidis, B.~Sundborg, and R.~van Unge, ``{Hamiltonian BRST quantization of the conformal string},'' {\em Nucl. Phys. B} {\bf 440} (1995) 495--520, \href{http://www.arXiv.org/abs/hep-th/9410143}{{\tt hep-th/9410143}}.

\bibitem{Jensen:1996dc}
B.~Jensen and U.~Lindstrom, ``{Classical interactions for tensionless strings},'' {\em Phys. Lett. B} {\bf 398} (1997) 83--87, \href{http://www.arXiv.org/abs/hep-th/9612213}{{\tt hep-th/9612213}}.

\bibitem{Bagchi:2020fpr}
A.~Bagchi, A.~Banerjee, S.~Chakrabortty, S.~Dutta, and P.~Parekh, ``{A tale of three \textemdash{} tensionless strings and vacuum structure},'' {\em JHEP} {\bf 04} (2020) 061, \href{http://www.arXiv.org/abs/2001.00354}{{\tt 2001.00354}}.

\bibitem{Bagchi:2021rfw}
A.~Bagchi, M.~Mandlik, and P.~Sharma, ``{Tensionless tales: vacua and critical dimensions},'' {\em JHEP} {\bf 08} (2021) 054, \href{http://www.arXiv.org/abs/2105.09682}{{\tt 2105.09682}}.

\bibitem{Bagchi:2019cay}
A.~Bagchi, A.~Banerjee, and P.~Parekh, ``{Tensionless Path from Closed to Open Strings},'' {\em Phys. Rev. Lett.} {\bf 123} (2019), no.~11, 111601, \href{http://www.arXiv.org/abs/1905.11732}{{\tt 1905.11732}}.

\bibitem{Bagchi:2022iqb}
A.~Bagchi, D.~Grumiller, and M.~M. Sheikh-Jabbari, ``{Horizon strings as 3D black hole microstates},'' {\em SciPost Phys.} {\bf 15} (2023), no.~5, 210, \href{http://www.arXiv.org/abs/2210.10794}{{\tt 2210.10794}}.

\bibitem{Banerjee:2024fbi}
A.~Banerjee, R.~Chatterjee, and P.~Pandit, ``{Tensionless strings in a Kalb-Ramond background},'' {\em JHEP} {\bf 06} (2024) 067, \href{http://www.arXiv.org/abs/2404.01385}{{\tt 2404.01385}}.

\bibitem{Figueroa-OFarrill:2025njv}
J.~Figueroa-O'Farrill, E.~Have, and N.~A. Obers, ``{Quantum carrollian bosonic strings},'' \href{http://www.arXiv.org/abs/2509.04397}{{\tt 2509.04397}}.

\bibitem{Hagedorn:1965st}
R.~Hagedorn, ``Statistical thermodynamics of strong interactions at high-energies,'' {\em Nuovo Cim. Suppl.} {\bf 3} (1965)
147.

\bibitem{Lizzi:1986nv}
F.~Lizzi, B.~Rai, G.~Sparano, and A.~Srivastava, ``{Quantization of the Null String and Absence of Critical Dimensions},'' {\em Phys. Lett. B} {\bf 182} (1986) 326--330.

\bibitem{Bagchi:2021ban}
A.~Bagchi, A.~Banerjee, S.~Chakrabortty, and R.~Chatterjee, ``{A Rindler road to Carrollian worldsheets},'' {\em JHEP} {\bf 04} (2022) 082, \href{http://www.arXiv.org/abs/2111.01172}{{\tt 2111.01172}}.

\bibitem{Bagchi:2023cfp}
A.~Bagchi, A.~Banerjee, J.~Hartong, E.~Have, K.~S. Kolekar, and M.~Mandlik, ``{Strings near black holes are Carrollian},'' {\em Phys. Rev. D} {\bf 110} (2024), no.~8, 086009, \href{http://www.arXiv.org/abs/2312.14240}{{\tt 2312.14240}}.

\bibitem{Bagchi:2024rje}
A.~Bagchi, A.~Banerjee, J.~Hartong, E.~Have, and K.~S. Kolekar, ``{Strings near black holes are Carrollian. Part II},'' {\em JHEP} {\bf 11} (2024) 024, \href{http://www.arXiv.org/abs/2407.12911}{{\tt 2407.12911}}.

\bibitem{Sheikh-Jabbari:2026tpf}
M.~M. Sheikh-Jabbari and H.~Yavartanoo, ``{Null Strings Gauged and Reloaded, II: Consistent Classical Treatment of the Null Strings},'' \href{http://www.arXiv.org/abs/2605.26822}{{\tt 2605.26822}}.

\bibitem{Lindstrom:2026quz}
U.~Lindstr{\"o}m, ``{Symmetries of tensionless strings},'' \href{http://www.arXiv.org/abs/2605.26185}{{\tt 2605.26185}}.

\bibitem{Duary:2026rlo}
S.~Duary and S.~Maji, ``{Path integral quantization of tensionless bosonic strings with Carroll-Weyl ghosts},'' \href{http://www.arXiv.org/abs/2606.04999}{{\tt 2606.04999}}.

\bibitem{Lindstrom:2026zno}
U.~Lindstr{\"o}m, ``{The conformal null string in $d+2$ and $d$ dimensions},'' \href{http://www.arXiv.org/abs/2606.22498}{{\tt 2606.22498}}.

\bibitem{Green:1987sp}
M.~B. Green, J.~H. Schwarz, and E.~Witten, {\em Superstring Theory}.
\newblock Cambridge University Press, 1987.
\newblock Vol. 1: {I}ntroduction.

\bibitem{Polchinski:1998rq}
J.~Polchinski, {\em String theory}.
\newblock Cambridge University Press, 1998.
\newblock Vol. 1: {A}n Introduction to the Bosonic String.

\bibitem{Hopkinson:1975pm}
J.~F.~L. Hopkinson, R.~W. Tucker, and P.~A. Collins, ``{Quantum Strings and the Functional Calculus},'' {\em Phys. Rev. D} {\bf 12} (1975) 1653.

\bibitem{Kanatchikov:2000yh}
I.~V. Kanatchikov, ``{Precanonical quantization and the Schrodinger wave functional},'' {\em Phys. Lett. A} {\bf 283} (2001) 25--36, \href{http://www.arXiv.org/abs/hep-th/0012084}{{\tt hep-th/0012084}}.

\bibitem{Hatfield:2019sox}
B.~Hatfield, {\em {Quantum Field Theory Of Point Particles And Strings}}.
\newblock CRC Press, 5, 2019.

\bibitem{Murase:2015yaa}
K.~Murase, ``{Anomaly of Tensionless String in Light-cone Gauge},'' \href{http://www.arXiv.org/abs/1503.01450}{{\tt 1503.01450}}.

\bibitem{Chen:2026klv}
B.~Chen and Z.~Hu, ``{Symmetries and Critical Dimensions of Tensionless Branes},'' \href{http://www.arXiv.org/abs/2604.01883}{{\tt 2604.01883}}.

\bibitem{Gradshteyn-Ryzhik}
I.~S. Gradshteyn and I.~M. Ryzhik, {\em Table of integrals, series, and products}.
\newblock Elsevier/Academic Press, Amsterdam, seventh~ed., 2007.
\newblock Translated from the Russian, Translation edited and with a preface by Alan Jeffrey and Daniel Zwillinger, With one CD-ROM (Windows, Macintosh and UNIX).

\bibitem{Bonelli:2003kh}
G.~Bonelli, ``{On the tensionless limit of bosonic strings, infinite symmetries and higher spins},'' {\em Nucl. Phys. B} {\bf 669} (2003) 159--172, \href{http://www.arXiv.org/abs/hep-th/0305155}{{\tt hep-th/0305155}}.

\bibitem{Sundborg:2000wp}
B.~Sundborg, ``{Stringy gravity, interacting tensionless strings and massless higher spins},'' {\em Nucl. Phys. B Proc. Suppl.} {\bf 102} (2001) 113--119, \href{http://www.arXiv.org/abs/hep-th/0103247}{{\tt hep-th/0103247}}.

\bibitem{Fotopoulos:2010ay}
A.~Fotopoulos and M.~Tsulaia, ``{On the Tensionless Limit of String theory, Off - Shell Higher Spin Interaction Vertices and BCFW Recursion Relations},'' {\em JHEP} {\bf 11} (2010) 086, \href{http://www.arXiv.org/abs/1009.0727}{{\tt 1009.0727}}.

\end{thebibliography}
\end{document}